\documentclass[12pt,titlepage]{article}
\usepackage{amsmath}
\usepackage{comment}
\usepackage{amssymb}
\usepackage{graphicx}
\usepackage{caption2}
\usepackage{amsfonts}
\usepackage{cite}
\usepackage{lineno}
\usepackage{color}
\usepackage{ulem}
\usepackage{booktabs}
\usepackage{nomencl}
\usepackage{soul}
\makenomenclature

\newcommand{\be}{\begin{equation}}
\newcommand{\ee}{\end{equation}}
\newcommand{\bea}{\begin{eqnarray}}
\newcommand{\eea}{\end{eqnarray}}
\newcommand{\bef}{\begin{figure}}
\newcommand{\ef}{\end{figure}}
\newcommand{\bt}{\begin{tabular}}
\newcommand{\et}{\end{tabular}}
\newcommand{\bno}{\begin{enumerate}}
\newcommand{\eno}{\end{enumerate}}

\def\3{\ss}

\catcode`\"=\active
\def"{\accent'177}

\begin{document}

\begin{center}

{\Large A physical perturbation based study on the prediction of free-fall disks with chaotic modes in the water} 

\vspace{0.3cm}
Tianzhuang Xu $^{1,2}$ 
Bo Zhang $^{1,2}$
Jing Li $^{1,2}$\footnote{Corresponding author, Email address: lijing\_@sjtu.edu.cn} \\Zhihui Li$^{4,5}$ and Shijun Liao $^{1,2,3}$ 

\vspace{0.3cm}
$^1$ Marine Numerical Experiment Center, State Key Laboratory of Ocean Engineering, Shanghai 200240, China

$^2$ School of Naval Architecture, Ocean and Civil Engineering, Shanghai Jiao Tong University, Shanghai 200240, China

$^3$ School of Physics and Astronomy, Shanghai Jiao Tong University, Shanghai 200240, China

$^4$  China Aerodynamics Research and Development Center, P.O.Box 211, Mianyang 621000, China

$^5$ National Laboratory for Computational Fluid Dynamics, BUAA, Beijing 100191, China
 \end{center}

\hspace{-0.85cm} {\bf Abstract}  { 
We report a phenomenon that physical perturbations sometimes can benefit the certainty of a free-fall motion with chaotic modes,
albeit, as commonly believed, they can ruin it.
We statistically compare those factors that may lead to uncertainty, 
by which we find that the growth of the standard deviation of the landing locations is directly determined by
the physical perturbations.
A significant yardstick is defined in the meantime.
This temporal criterion is of big relevance to the replicability of such problems experimentally,
although they are inherently chaotic.
Our hypothesis is verified by experiments from other literature.
This outcome also provides a practical strategy to evaluate the credible prediction time by estimating the disturbances from physical parameters as a priori.
}

\hspace{-0.85cm} {\bf Key words}  numerical simulation; free-fall; parametric noises; chaos

\hspace{-0.85cm} {\bf AMS Subject Classifications}   37D45,  34C28,  65P20, 65M12

\nomenclature{$x, y$}{Horizontal and vertical location of the falling disk in global reference system}
\nomenclature{$\theta$}{Angle of attack}
\nomenclature{${V}_{x^{\prime}}, {V}_{y^{\prime}}$}{Speed of the falling disk in local $x^{\prime}$ and $y^{\prime}$ direction }
\nomenclature{$F_{x^{\prime}}, F_{y^{\prime}}$}{Viscous drag of the falling disk in local $x^{\prime}$ and $y^{\prime}$ direction}
\nomenclature{$\tau$}{Viscous torque of the falling disk}
\nomenclature{$I^{*}$}{Dimensionless moment of inertia} 
\nomenclature{$\Gamma$}{Circulation term of the falling disk resulted from the translational and rotational movement}
\nomenclature{$C_R, C_T$}{Circulation coefficients of rotation and translation} 
\nomenclature{$A, B$}{Viscous drag coefficients, the overall and the angle of attack effects} 
\nomenclature{$\mu_1, \mu_2$}{Viscous torque coefficients, linear and quadratic part} 
\nomenclature{$\delta$}{Threshold value of the convergence check to determine the predicable time $T_p$}
\nomenclature{$\Delta$}{Prefix denoting the absolute errors of parameters and locations}
\nomenclature{$\epsilon$}{Relative physical perturbations}
\nomenclature{$T_p$}{Predicable time of a single trajectory with physical perturbations}
\nomenclature{$T_{ti}$}{Predicable time (temporal invariant model)}
\nomenclature{$T_{tv}$}{Predicable time (temporal variant model)}
\nomenclature{$D_{\max}$}{The maximum absolute difference between the cumulative distribution}
\nomenclature{$h$}{Time step of ODE integrators}
\nomenclature{$k_i$}{Coefficients of the right-hand side of temporal discretization}
\nomenclature{$K$}{Decimal digits to store data}
\nomenclature{$\mathcal{O}\left(h^{N}\right)$}{The order of truncation errors}
\nomenclature{$\rho$}{Pearson correlation coefficients of the error function}
\nomenclature{$\sigma$}{The standard deviation of variables}
\printnomenclature

\section{Introduction}

Objects falling freely in fluids widely exist, and some of them hold a tight connection to industries
or human lives.
The free-fall phenomena are simple physically with only gravity and fluid dynamics on it,
but they inherit different falling modes containing complex patterns.
Dated back to the nineteenth century, Maxwell already noticed two modes: tumbling and fluttering\cite{maxwell_particular_1854}, but no one knew how to distinguish them.
Willmarth's series of experiments quantitatively deepens the understanding of the problem\cite{willmarth_steady_1964}. 
He concluded that the falling modes only depend on the dimensionless moment of inertia and the Reynolds number when the ratio of the disk thickness to the diameter is small. 
Based on Willmarth's methodology, Field et al. further confirmed that chaotic modes also exit\cite{field_chaotic_1997}, and free-fall disk problem is actually a chaotic system.
With the help of the recent advances in experimental technologies, more complex and delicate falling patterns were discovered in three dimensions (3D).
By Zhong et al. 's study how two-dimensional fluttering modes transform to three-dimensional spiral modes was illustrated\cite{zhong_experimental_2013, lee_experimental_2013}.
Based on these spiral modes, Kim et al. further found a mutative falling mode with helical and conical motions when two disks are rigidly linked \cite{kim_free-fall_2018}.
Although these three-dimensional structures may look special, given the temporal evolution of the angle of attack, they are still not beyond the scope of the generalized fluttering modes.
So far, the stable, fluttering, tumbling, and chaotic modes are the four elementary falling modes which cover nearly every known falling pattern, even for objects with complex shapes, such as cylinders, polygons, cones, and particles\cite{auguste_falling_2013,chrust_numerical_2013,toupoint_kinematics_2019,amin_role_2019,kim_free_2020,esteban_edge_2018,esteban_three_2019}.

In addition to the fluid dynamic parameters and dynamic properties of the disk, 
many other factors can influence the free-fall motions.
Lee et al. found that the bristled structure of disks could strengthen the stability of falling behavior\cite{lee_stabilized_2020},
while the eccentricity can bring risk.
Zhou et al. reported that a small eccentricity can make the chaotic falling happen\cite{zhou_eccentric_2021}.
This conclusion is aligned with the analytical theory of re-entry capsule derived by Aslanov et al. \cite{aslanov_chaotic_2016}.
Besides, despite the stable falling feature of the conical shape\cite{amin_role_2019}, the small eccentricity can also result in chaos pointed out by Aslanov et al.
All the mentioned observations by researchers have shown that 
physical parameters consist of flow states, dynamic properties and geometries of the disk, etc., can more or less decide the falling behaviors.
In the real circumstances, the precision of these parameters is hardly controlled.
They are constantly accompanied by perturbations and noises.
Look at the turbulent flow full of fluctuations for an example,
and the trajectories of disks will be more likely fluttering chaotic \cite{esteban_disks_2020}.

For practical purposes, people prefer to choose numerical simulation to predict
the landing positions of falling objects.
The drift simulation of MH370 debris \cite{nhess-16-1623-2016}, for example, is a necessary approach to help
the rescue teams to locate the search area.
However, there is a large proportion of them falling down to the bottom of the sea, after drifting with sea flows.
An accurate prediction will definitely save time and money, while, as mentioned above, the falling procedure is constantly accompanied with noises.
Therefore, the motivation of this article is to find a practical strategy
to tell people to what time extent the prediction will be valid without any posteriori data.
The equivalent statement is that the reproducibility problem will not occur before the time criterion is reached
experimentally.
Since in laboratory test, a delicate experimental result with chaotic modes can only be replicated
probably because the noises have not been amplified to an apparent level.
We focus on a simplified analogue, free-fall disks in the water,
to be specific, holding trajectories with chaotic modes given the sensitivity to the environment, just like the literature said that ``the apparently chaotic motion is radically different from the periodic fluttering and tumbling ... is much more sensitive to experimental noise than in the periodic regions'' \cite{andersen_unsteady_2005}.

Here, we use a quasi-static model, the Andersen-Pesavento-Wang (APW) model, based on experiments.
``Quasi-static'' means that the reaction of the fluid on objects is decoupled from the fluid-structure interactions but governed by additional models only dependent on the kinematic states of objects.
The six degree-of-freedom (DoF) orbit model is a typical example of quasi-static models where drag, lift, and torque are computed by the product of fluid dynamic coefficients consisting of the Euler angle of objects, the square of the velocity, and the atmospheric density\cite{ronse_statistical_2014}.
Compared with the six DoF orbit model, free-fall models of disks in the water are reinforced with additional corrections to improve the precision.
The APW model is regarded as the most elaborate quasi-static model among them\cite{andersen_unsteady_2005, andersen_analysis_2005}.
Despite all this, it was reported that the reliable~(convergent) long-term prediction of chaotic motion 
without any physical perturbations is inaccessible because of the numerical errors\cite{xu_accurate_2021}.
Fortunately, this imperceptible factor can be expelled via the Clean Numerical Simulation (CNS), 
an arbitrary precision numerical scheme proposed by Liao \cite{liao_reliability_2008,liao_clean_2017 ,xu_accurate_2021}.
By this means, we shall study those noises which exist in the real world without worry about the contamination from numerical errors.

What are the physical perturbations? 
Though the derivations of the quasi-static models are mathematically strict based on 
reasonable assumptions of fluid mechanics, the determination of parameters in such models is not reliable.
A few of them cannot even be expressed explicitly.
Thus, they can only be determined via data-fitting from additional direct numerical simulation or experimental results, 
which inevitably introduces fitting errors or measurement errors.
The measurement errors in particular include fluctuations and perturbations of fields 
during the flow dynamic procedure, and other inevitable errors between the approximation and the physics.
Hence, we name them physical perturbations or physical perturbations.

This paper is organized as follows.
The APW model and the CNS method are shortly summarized in the next section.
In the third section,
the failure of predicting a single trajectory containing physical perturbations is then demonstrated. 
After that, we conduct Monte-Carlo simulations to explain the transportation of physical perturbations
in the aspect of the temporal evolution, 
since the probability distribution is the information of interest in most cases.
Then, the main results are given by bringing in noises changing over time, and some comparisons are made, 
which is followed by the fourth section where some necessary discussions on the experimental replicability are carried out. 

\section{Mathematical model and numerical algorithm}

\subsection{Andersen-Pesavento-Wang (APW) model}

\begin{figure}[!t]
\centering
\includegraphics[scale=1]{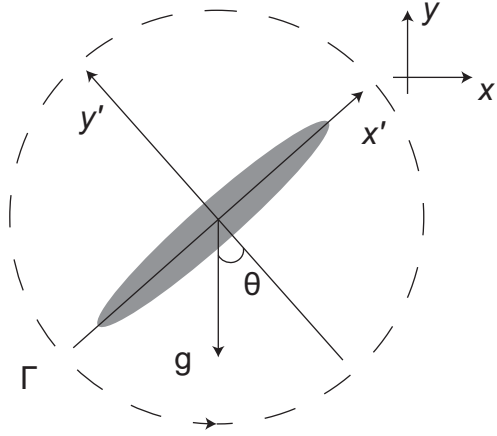}
\caption{The local reference frame $(x^{\prime}, y^{\prime})$ fixed with the disk and the global reference frame $(x,y)$ of a freely falling two-dimensional desk, where $\theta$ denotes the rotation angle of the disk, $g$ is the acceleration due to gravity,  $\Gamma$ is the circulation, respectively. }
\label{fig1}
\end{figure}

The APW model is a quasi-static model describing the free-fall motion of disks in the water, strictly derived from the Navier--Stokes equation under the assumption of ideal fluids and amended with additional lift, drag, and torque models\cite{andersen_analysis_2005, andersen_unsteady_2005}. Its dimensionless form in the case of thin disks is written as follows
\begin{equation}
\left\{
\begin{aligned}
I^{*} \dot{V}_{x^{\prime}} &=\left(I^{*}+1\right) \dot{\theta} V_{y^{\prime}}-\Gamma V_{y^{\prime}}-\sin \theta-F_{x^{\prime}}, \\
\left(I^{*}+1\right) \dot{V}_{y^{\prime}} &=-I^{*} \dot{\theta} V_{x^{\prime}}+\Gamma V_{x^{\prime}}-\cos \theta-F_{y^{\prime}}, \\
\frac{1}{4}\left(I^{*}+\frac{1}{2}\right) \ddot{\theta} &=-V_{x^{\prime}} V_{y^{\prime}}-\tau,
\end{aligned}
\right.
\label{apwmodel}
\end{equation}
with the coordinate transformation, and the coordinate is demonstrated in Fig. \ref{fig1}
\begin{equation}
\left\{
\begin{aligned}
\dot{x}& =V_{x^{\prime}} \cos \theta-V_{y^{\prime}} \sin \theta, \\
\dot{y}& =V_{x^{\prime}} \sin \theta+V_{y^{\prime}} \cos \theta. \\
\end{aligned}
\right.
\end{equation}
Andersen et al. proposed the implementation of a novel lift model closed by the velocities and angular velocities plus the Joukovsky theorem to introduce the lift and torque amendment
\begin{equation}
\Gamma=\frac{2}{\pi}\left(-\frac{C_{T}  V_{x^{\prime}} V_{y^{\prime}}}{\sqrt{V_{x^{\prime}}^{2}+V_{y^{\prime}}^{2}}}+C_{R} \dot{\theta}\right).
\end{equation}
The drag and dissipative torque terms are in a classic quadratic form
\begin{equation}
\left( F_{x^{\prime}}, F_{y^{\prime}}\right) =\frac{1}{\pi}\left(A-B \frac{V_{x^{\prime}}^{2}-V_{y^{\prime}}^{2}}{V_{x^{\prime}}^{2}+V_{y^{\prime}}^{2}}\right) \sqrt{V_{x^{\prime}}^{2}+V_{y^{\prime}}^{2}}\left(V_{x^{\prime}}, V_{y^{\prime}}\right), \hspace{0.5cm} \tau=\left( \mu_{1}+\mu_{2}|\dot{\theta}|\right) \dot{\theta}.
\label{abs}
\end{equation}
This model has seven dimensionless parameters $I^*, C_T, C_R, A, B, \mu_1$, and $\mu_2$. 
For the seven parameters, only the dimensionless moment of inertia $I^*$ is explicitly expressed, and determined by the geometry of disks and the density ratio of the disk to the water. 
Other coefficients $C_T, C_R, A, B, \mu_1$, and $\mu_2$ cannot be obtained by theoretical reduction,
but by fitting the measured data.
That is where physical perturbations inevitably come from.
We use the prefix $\Delta$ to represent them. 
For example, $\Delta \mu_2$ represents the noises of $\mu_2$.  
For convenience, we define the relative ratio of the parameter noise to the true value of the parameter,
$\epsilon = \frac{\Delta P}{P}$, where $P$ stands for an arbitrary parameter, 
which will provide a direct measurement of how large the magnitude of the noise would be.

\subsection{Clean Numerical Simulation (CNS)}
Generally speaking, CNS is a series of strategies with a purpose to control ``numerical noises'' arbitrarily. 
Its main point is decreasing the truncation error and the round-off error to a required level. 

Truncation errors come from the discretization of continuous systems. 
Numerical methods have the following general form
\begin{equation}
	f(t+h)=f(t)+h\cdot RHS(t).
	\label{eq:error}
\end{equation}
where $RHS(t)$ denotes the right-hand side. 
It varies according to the numerical methods. 
For the N-th order Runge-Kutta method family, N-step multi-step method, and N-th order Taylor series method, the right-hand side may read
\begin{equation}
	\left\{\begin{array}{l}
	R H S(t)=\sum\limits_{i=1}^{N} k_{i} f\left(t_{k_{i}}\right)+\mathcal{O}\left(h^{N}\right), \\
	R H S(t)=\sum\limits_{i=1}^{N} k_{i} f\left(t_{i-N-1}\right)+\mathcal{O}\left(h^{N}\right), \\
	R H S(t)=\sum\limits_{i=1}^{N} \dfrac{f^{(i)}(t)}{i !} h^{i-1}+\mathcal{O}\left(h^{N}\right),
	\end{array}\right.
\end{equation}
where $\mathcal{O}(h^N)$ is the order of the global truncation errors. 
The CNS suggests that reducing the timestep $h$ or increasing $N$ with high order methods 
can decrease truncation errors to a small level which will not damage the long-term prediction. 
With the help of extended precision formats to store extra digits in computers, 
round-off errors can be largely avoided. 
Consequently, the numerical noises of any simulation can be controlled arbitrarily small. 
The detailed numerical scheme used in the present study is referred to the previous article \cite{xu_accurate_2021}.

\section{Chaotic falling containing physical perturbations}
As mentioned in the first section, free-fall disks will lead to main four categories of motions
until they touch the bottom.
In this manuscript, only chaotic mode is the primary objective for its sensitive nature to 
physical perturbations.
This section will focus on the impact of these physical perturbations on the prediction of the chaotic falling
by the CNS.

\subsection{Single trajectory}

\begin{figure}[!t]
\centering
\includegraphics[scale=.4]{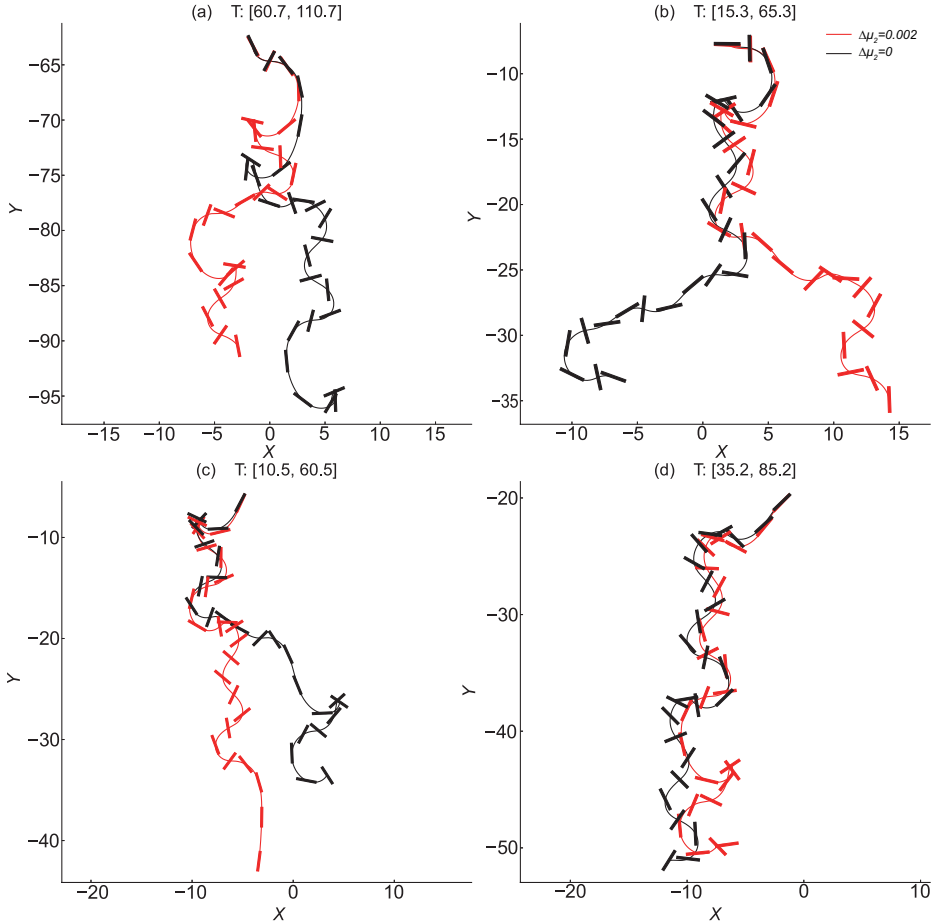}
\caption{Detail comparison between trajectories with and without physical perturbations, which is colored by red and black. (a) (b) (c) (d) correspond to case 1 to 4 respectively.}
\label{fig2}
\end{figure}

First, the single trajectory is studied with $I^* = 2.2$, where the falling is chaotic.
To have an impression of the effects of physical perturbations on a single trajectory, let us consider there are one percent physical perturbations $\epsilon = 1\%$ on the quadratic dissipative torque coefficients $\mu_2=0.2$, that is $\Delta\mu_2 = \epsilon \mu_2 = 0.002$.
The four initial conditions studied in the previous work\cite{xu_accurate_2021} are chosen first. $(\theta, V_{x^{\prime}}, V_{y^{\prime}}, \dot{\theta})$ for different cases is $(0, 0, 0.01, 0)$, $(1, 0, 0.01, 0)$, $(0, 0, 0.01, 1)$, and $(1, 0, 0.01, 1)$, respectively. They correspond to the initial conditions of falling without initial angle of attack and angular velocity, only with initial angle of attack, only with initial angular velocity and with both initial angle of attack and angular velocity.
The computational results are demonstrated in Fig.~\ref{fig2}, visualizing the separations of plates due to physical perturbations.
The effects of physical perturbations are generally similar to numerical noises reported previously\cite{xu_accurate_2021}.
With minute physical perturbations as small as $1\%$, the computational results are totally ruined afterwards and lose their predictability.
Besides, since the scale of physical perturbations are way larger than numerical noises, the separations usually happen much earlier.
As demonstrated in Fig.~\ref{fig2}, the latest separation is in case 1 at about $50$ units of time (UT), 
which is far earlier than $200$ UT resulted from the numerical noise\cite{xu_accurate_2021},
because physical perturbations are usually far beyond the magnitude of numerical ones.
For example, if the plate rotates with an angular speed of $1$ rad/s
and with a parameter noise of $\epsilon=1\%$ for $\mu_2$, 
the error in the phase space can be amplified to a level as large as $10^{-3}$, 
while its counterpart is only $10^{-16}$ due to numerical noises.

To further quantify how long one could hold the predictability, we define a reliable prediction time $T_p$ to measure the duration of the simulation for a single trajectory 
until $|1-\frac{u_{p}} {u_{c}}|>\delta$ where $u_p$ and $u_c$ denote the locations of the plate with and without physical perturbations, respectively. 
Here we set the tolerance as $\delta=1 \%$ without loss of generality.

\begin{figure}[!t]
\centering
\includegraphics[scale=.5]{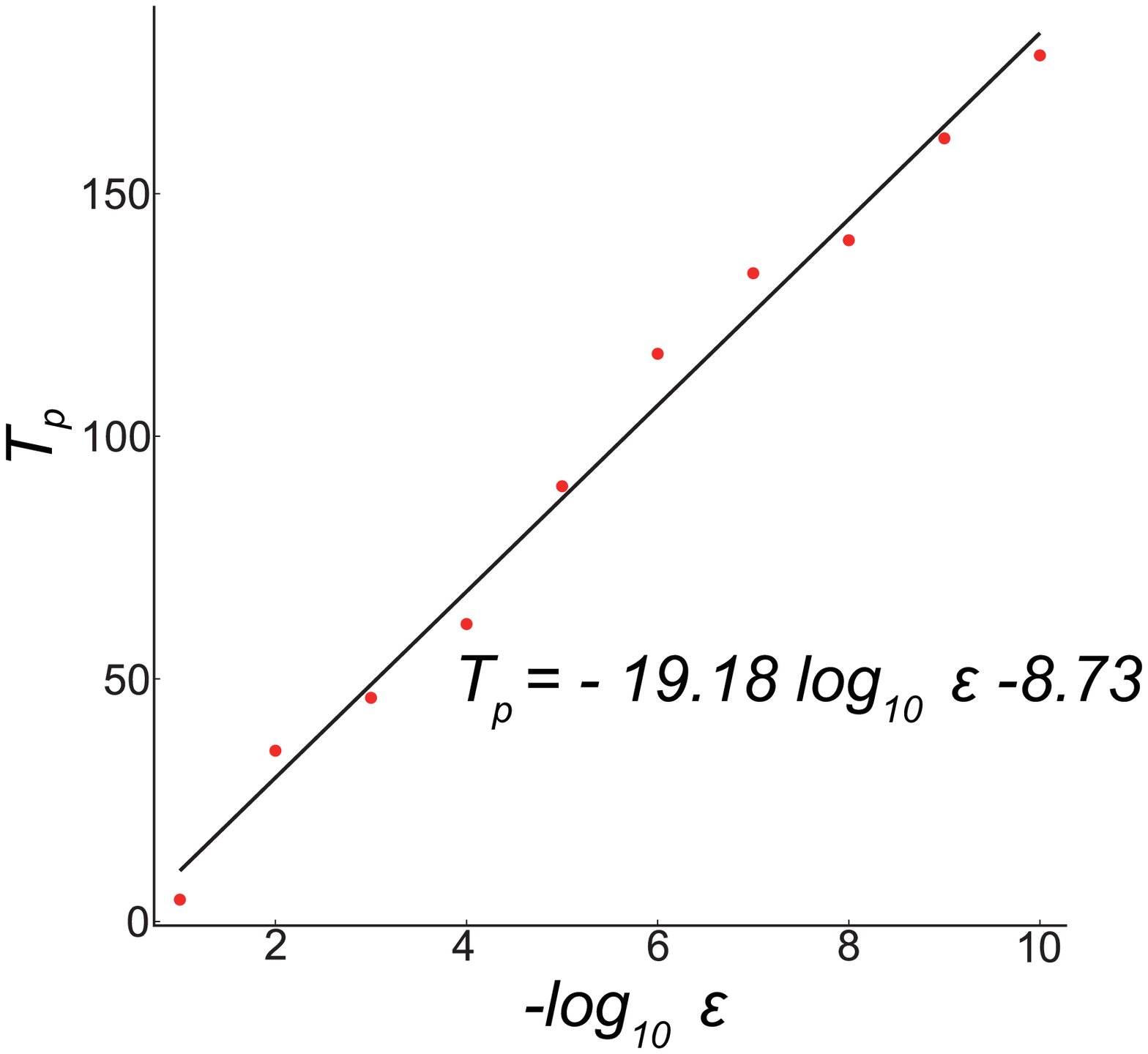}
\caption{The linear relationship between $T_p$ and the logarithm of $\epsilon$.}
\label{fig3}
\end{figure}

One example of $T_p$ is computed with case 4 and a series of varying physical perturbations $\epsilon$ of $\mu_2$.
The scaling law can be derived from the exponential growth of errors in chaotic systems\cite{liao_clean_2017}.
Practically, we could obtain the formula of $T_p$ with respect to $\epsilon$ with linear regressions to provides a rough estimation of the predictability.
The final fitted linear relationship of $T_p$ and $\epsilon$ is shown in Fig.~\ref{fig3}.

\begin{equation}
T_{p} \approx - 19.18 \log _{10} \epsilon - 8.73.
\label{eq:tp}
\end{equation}

Note that formula (\ref{eq:tp}) of the predicable time can only describe the relationship between relative parameter noise $\epsilon$ and $T_p$.
Nevertheless, $T_{p}$ also varies according to different initial conditions and parameters that contain noises.
For example, the divergence of case 1 happens significantly later than other cases.
As discussed qualitatively before, that is due to the initial condition of case 1 has longer transients before entering the chaotic state\cite{xu_accurate_2021}.
Also, different parameters have different physical meanings.
Thus, even with the same relative magnitude $\epsilon$, they can have quite different predicable times.
To further study the effects of initial conditions and parameters, we generate $100$ random initial conditions from the space $V_{x^\prime} \in (-3, 3), V_{x^\prime} \in (-1, 1), \theta \in (0, 2\pi), \dot{\theta} \in (-0.5\pi, 0.5\pi)$ uniformly, and consider the perturbed parameters $C_T, C_R, A, B, \mu_1, \mu_2, I^{*}$.

\begin{figure}[!t]
\centering
\includegraphics[scale=.4]{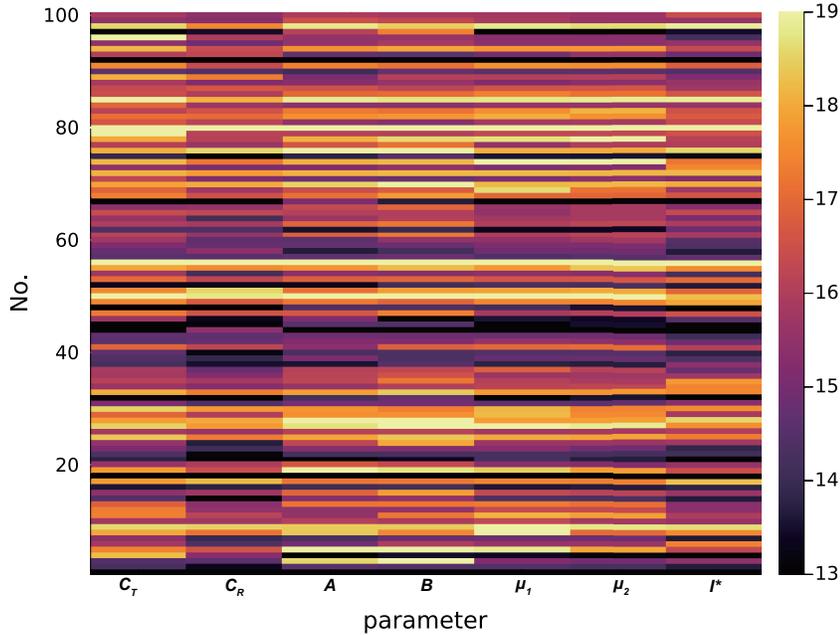}
\caption{The heat plot of the slope for the scaling law of $T_p$, where x axis corresponds to the perturbed parameters and the y axis corresponds to different initial conditions (where No. denotes the sequential numbers of different cases). The horizontal bar structure denotes the slope who is sensitive to initial conditions rather than the perturbed parameter.}
\label{fig4}
\end{figure}

From our numerical experiment, all the cases still apply to the reported scaling law.
Nevertheless, the slope and intercept vary by initial conditions and the parameters that contain noises.
Generally, in $90\%$ cases(from the fifth percentile to the ninety-fifth percentile), the computed slopes vary within the range $(13.0, 19.0)$.
As demonstrated in Fig.\ref{fig4}, the slope is sensitive to the initial condition, but not sensitive to which parameter is perturbed.
The intercept does not have specific laws, which is both sensitive to the initial condition and the parameter chosen to contain noises.
That can be explained from the perspective of dynamical system theory. 
The initial condition determines the manifold of the trajectories, that is no matter what noises are, they are amplified along with the manifold in a similar way.
Though different parameters can bring different noises, they are amplified similarly along with the propagation of trajectories.
The intercept, on the other hand, is closely related to the magnitude of physical perturbations, especially the noises introduced to trajectories at the first few steps.
Both initial conditions and parameters chosen are related to that.
For example, if the rotational dissipative torque term contains physical perturbations, it would bring much larger noises to trajectories if the initial condition has a large rotational velocity.

\subsection{Stochastic predictability}
As discussed, $T_p$ is a useful measurement for people to know straightforwardly when a single trajectory will diverge given exactly the value of the physical perturbation.
Nevertheless, such information is never available.
Practically, we can at most estimate the distributions of physical perturbations.
The methodology of estimating the predictability with physical perturbations that follows certain distributions is studied.

We consider two theoretical frameworks to introduce physical perturbations as random variables into the APW model.
The first is the temporal invariant model.
All random physical perturbations added are immutable all the time.
The second is the temporal variant model, where we consider a stochastic dynamical system that the parameters in the APW model fluctuate temporally around the true value by physical perturbations.

\begin{figure}[!t]
	\centering
	\includegraphics[scale=.6]{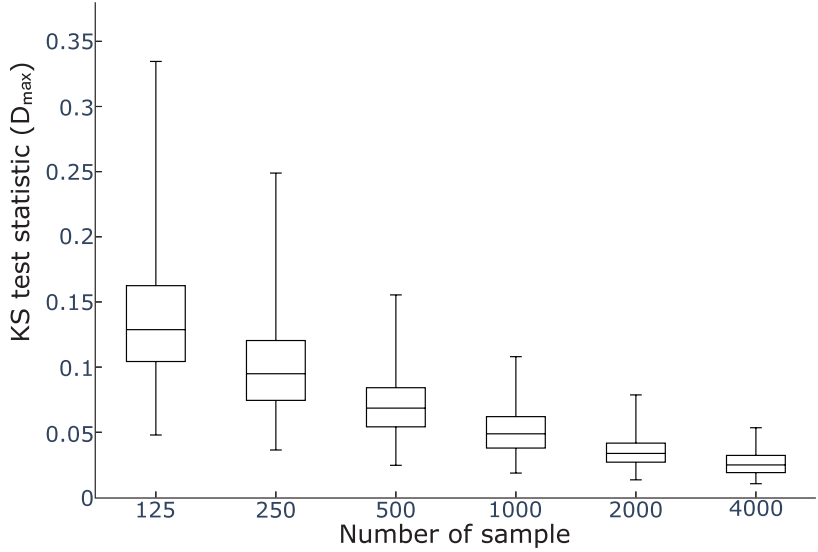}
	\caption{KS test statistic with number of sample}
	\label{fig5}
\end{figure}

To study the predictability under physical perturbations, Monte-Carlo simulations are used to access prediction intervals.
We made a convergence test with the two-sample Kolmogorov-Smirnov test (KS test) to determine the sampling number needed for normal distributions\cite{smith_uncertainty_2013}.
Here we firstly construct a sampling space $S$ with 10,000 samples to simulate the distribution at 200 UT. 
Then six sub-samples $S_{i} \in S(i=1,2,...,6)$  with different sizes $N_{i} = 125 \times 2^{i-1} (i=1,2,...,6)$ are defined, and sampled from the sampling space $S$ for $R_{i} = 125 \times 2^{6-i} (i=1,2,...,6)$ times. 
The statistic of the test $D_{\max}$ denotes the maximum absolute difference between the cumulative distribution functions (CDF):
\begin{equation}
    D_{\max }=\max _{x}(|\hat{F}(x)-F(x)|)
    \label{eq:df}
\end{equation}
where $\hat{F}$ and $F$ present the CDF of the sub-sample and the sample space, respectively.
As shown in Fig.\ref{fig5}, $D_{\max}$ decreases with the number of samples. 
Considering the trade-off between accuracy and computational efficiency, 1000 samples, with median error $4.8\%$ of the CDF, is enough for the representation of physical perturbations that follow normal distributions.

\subsubsection{Temporal invariant model}
Without loss of generality, we exemplify case 4 with $\mu_2$ containing physical perturbations. 
The physical perturbations are assumed to follow normal distributions according to the central limit theorem.
Suppose $\epsilon = 1\%$, and hence the mean is zero and the standard deviation $\sigma_{\Delta \mu_2}=0.002$ in the normal distribution.

For the purpose of prediction, the horizontal locations of the free-fall disks are of significant importance. 
We study the prediction errors of the horizontal location at each time step denoted as $\Delta X(t)$. 
The random variable includes information on the consequent prediction errors at a specific time.
$\Delta X(t)$ is studied from two distinct aspects. 
First, $\Delta X(t)$ can be seen as an error function of $\Delta \mu_2$ from a deterministic point of view.
As we will see, that function is a tool to visualize how physical perturbations hinder the long-term prediction.
Second, $\Delta X(t)$ is considered from the traditional perspective of statistics.
By analysis, the impact of physical perturbations on the prediction could be qualitatively categorized into three stages: determinism, transition, and randomness. 

\begin{figure}[!t]
\centering
\includegraphics[scale=.4]{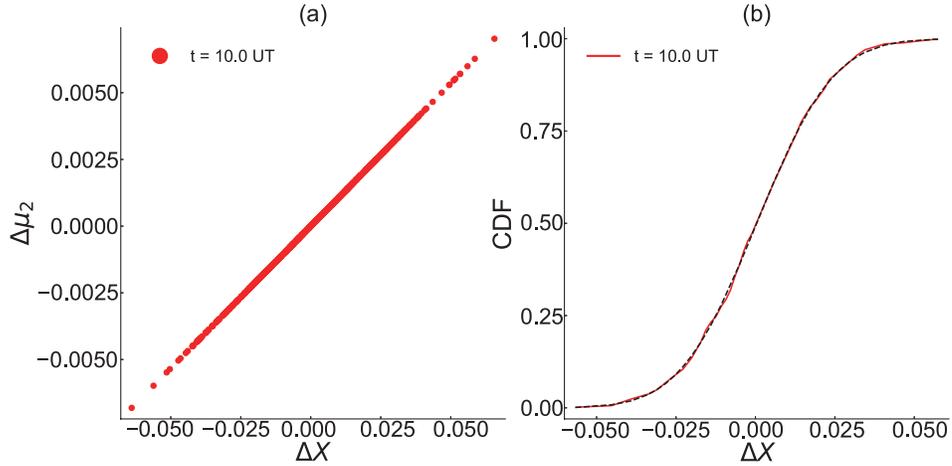}
\caption{The  example case of determinism stage at $10.0$ UT. (a) The error function; (b) red solid line: the CDF of $\Delta X$; black dashed line: the CDF of the corresponding normal distribution}
\label{fig6}
\end{figure}

At the first stage, the physical perturbations transfer into trajectories and result in a minute deviation.
As shown in Fig.\ref{fig6} (a), the error function is perfectly linear.
Due to the linearity, $\Delta X$ naturally applies to the same type of distributions as $\Delta \mu_2$ does.
The CDF of $X$ applies to a normal distribution.
During the stage, the shapes of both the error function and the CDF are stable.
All trajectories are ranked in order according to their physical perturbations.
In addition to that, the magnitude of $\Delta X$ is too small to harm the prediction.
Thus, any trajectory can be regarded as a satisfying prediction.
As to case 4, the determinism stage lasts until about $35$ UT.
This temporal scale of the determinism stage has a similar meaning as $T_p$ to some extent, but different in perspective in that it is a stochastic concept.

\begin{figure}[!t]
\centering
\includegraphics[scale=.4]{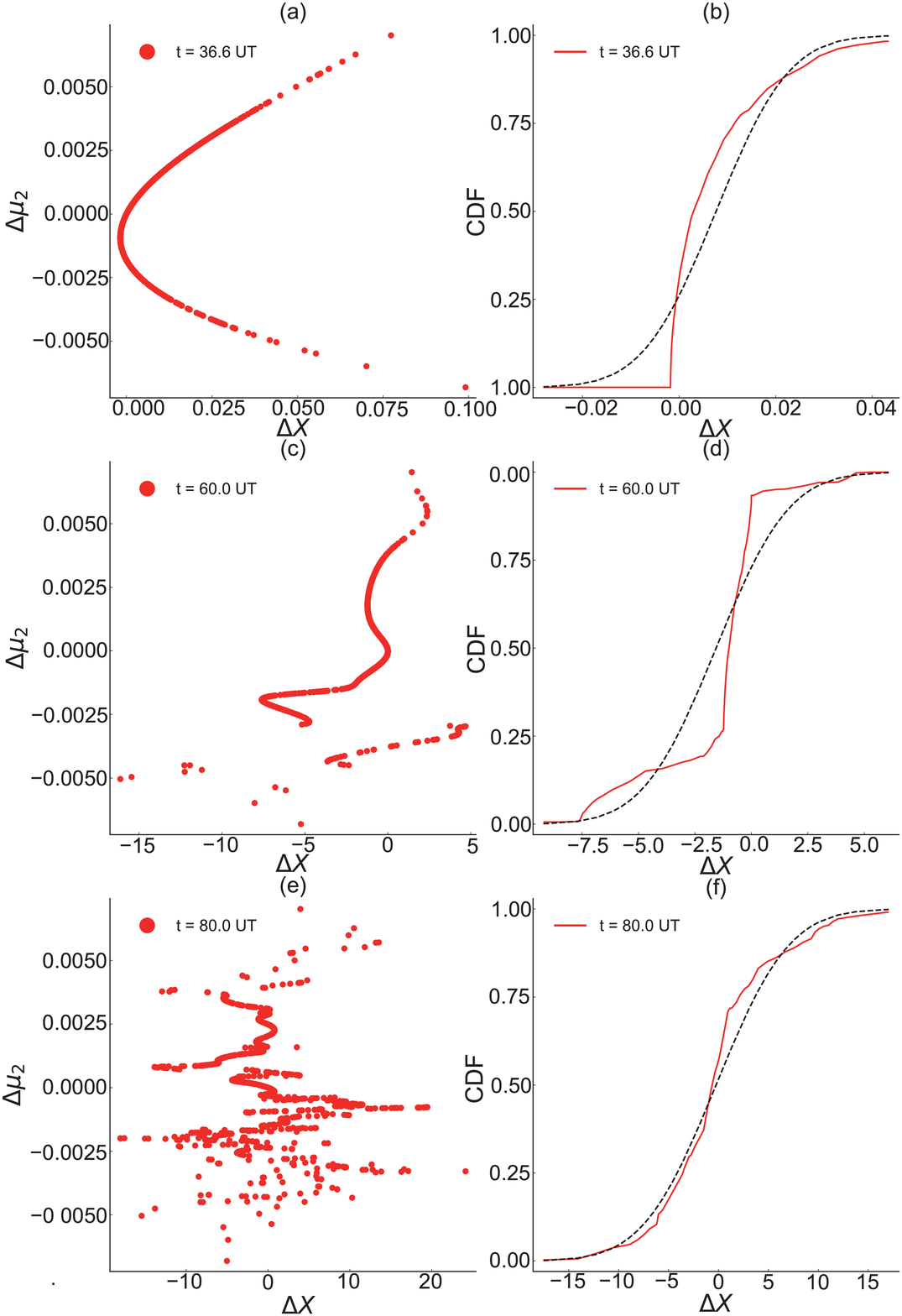}
\caption{The  example cases of transition stage at $36.6 UT, 60.0 UT, 80.0 UT$, respectively. (a)(c)(e) The error function; (b)(d)(f) red solid line: the CDF of $\Delta X$; black dashed line: the CDF of the corresponding normal distribution}
\label{fig7}
\end{figure}

Then comes the transition stage.  
Its features can be easily told from the plots of the error function and the CDF in Fig.\ref{fig7}.
During this stage, the spatial distribution of the disks is from order to disorder, and finally ends up with a
self-organized nature.
As to case 4, this process can be detailed as follows.
At around $36.6$ UT, the error function begins to lose its linearity.
The CDF, correspondingly, also distorts to the non-normal distribution.
After that, the shapes of both the error function and the CDF are continuously deforming.
Later at around $60.0$ UT, the continuity of the error function is partly destroyed.
The CDF thus develops into more complex one.
The error function keeps falling apart to stronger randomness while the CDF appears to turn back to the normal distribution at about $80.0$ UT.
At the moment, $\Delta X$ has grown to a macroscopically observable level,
affecting the reliable prediction of the motion of free-fall disks.
In addition to that, since both the error function and the CDF of $\Delta X$ keep intensely changing, 
it is challenging to even provide a stochastic prediction of the position of free-fall disks at this stage.

\begin{figure}[!t]
\centering
\includegraphics[scale=.4]{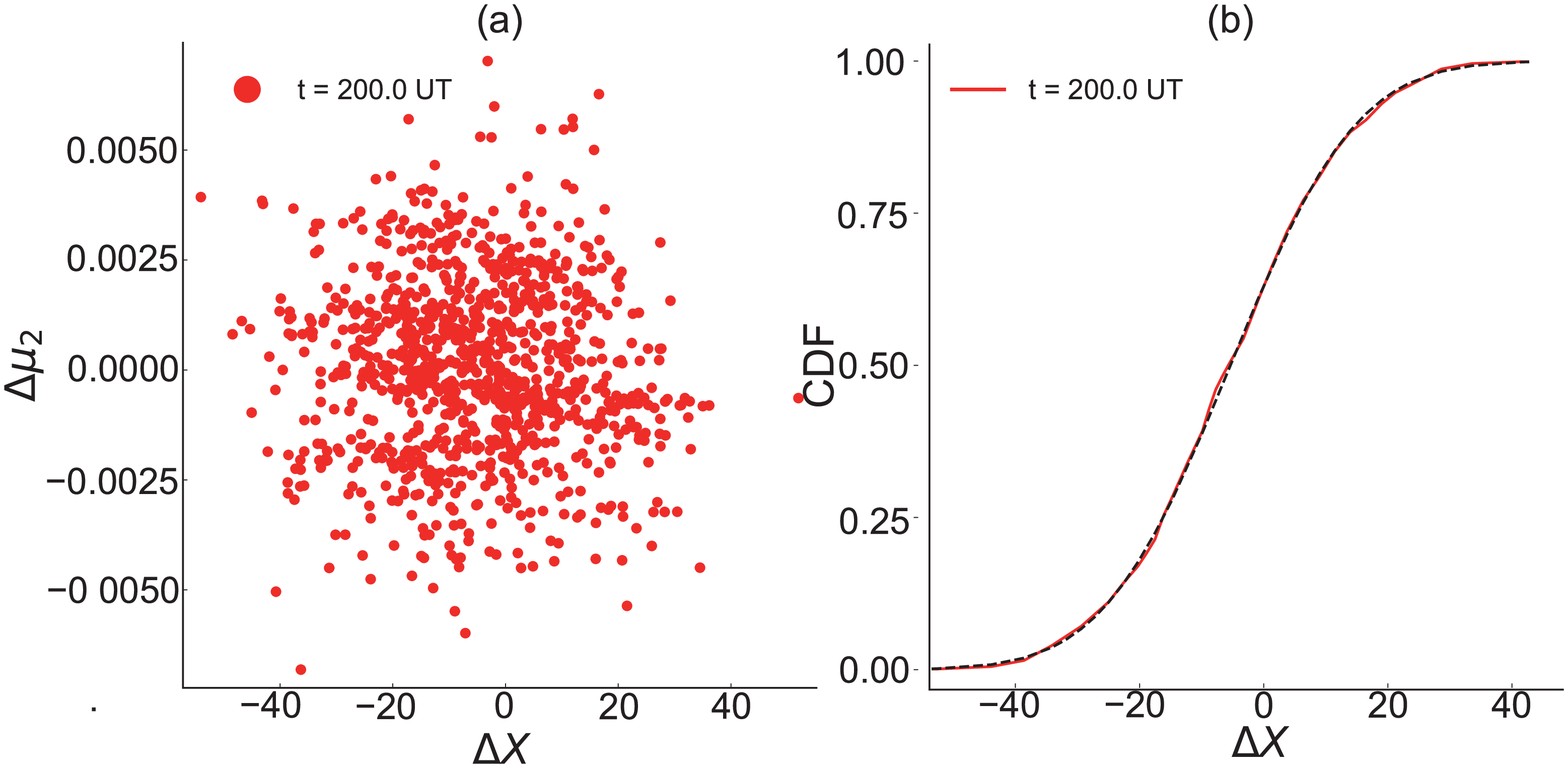}
\caption{The example case of randomness stage at $200 UT$. (a) The error function; (b) red solid line: the CDF of $\Delta X$; black dashed line: the CDF of the corresponding normal distribution}
\label{fig8}
\end{figure}

The statistic characteristics return to another stable state after the transition stage as shown in Fig.\ref{fig8}.
The relationships between $\Delta X$ and $\Delta \mu_2$ are completely chaotic. 
Thus, we name it the randomness stage.
The CDF of $\Delta X$ comes back to the normal distribution.
This distribution promises that there is a stochastic prediction window at the final stage.
Nevertheless, it is worth noting that in the chaotic cases the magnitude of  $\Delta X$ keeps growing larger
and can be up to several hundred times larger than the disk width, 
which obviously provides no meaningful information regarding the falling trajectories. 
 
\begin{figure}[!t]
\centering
\includegraphics[scale=.4]{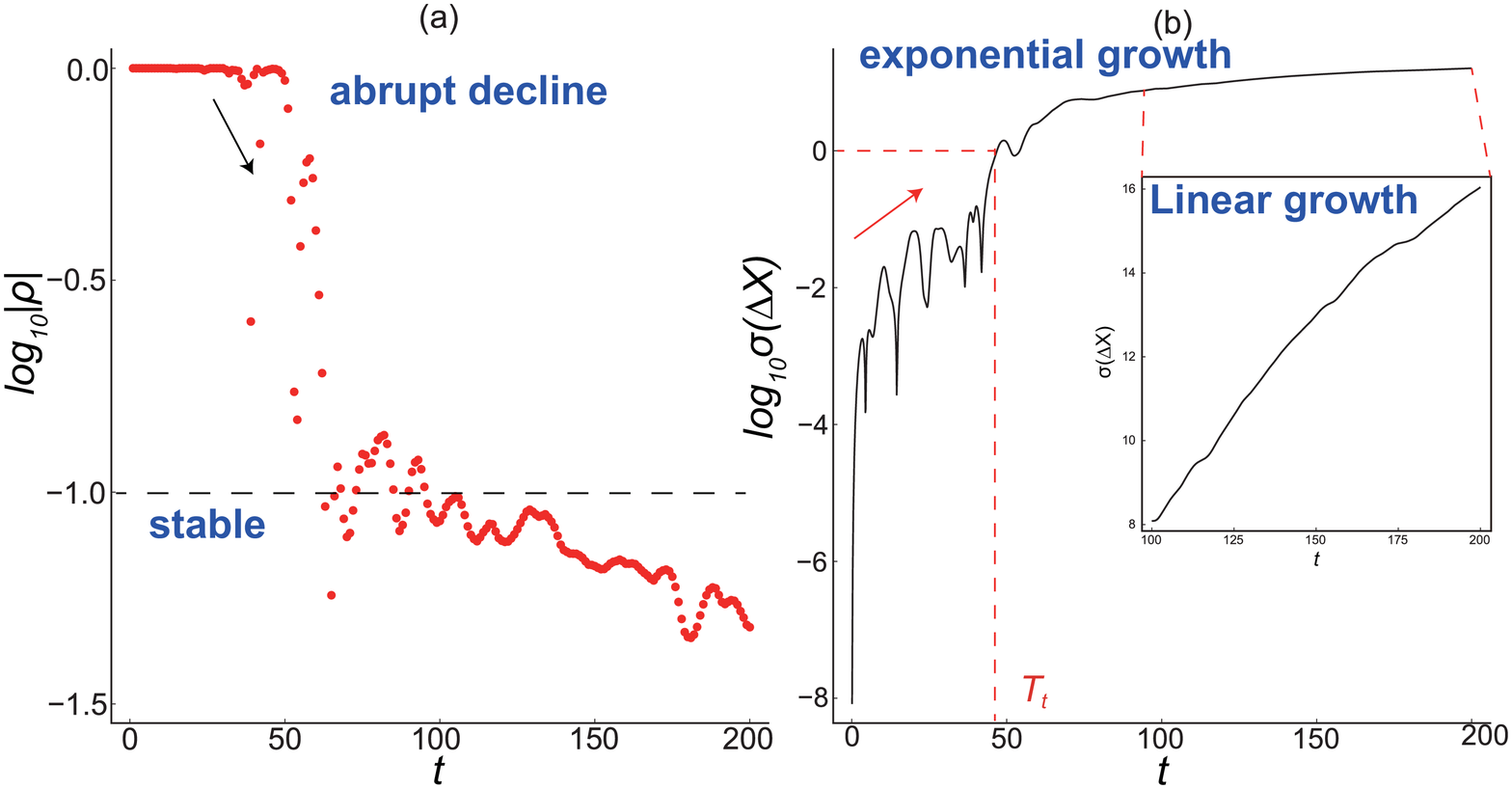}
\caption{(a) The temporal evolution of $\rho$; (b) The temporal evolution of $\sigma_{\Delta X}$.}
\label{fig9}
\end{figure}

To better understand the transition from determinism to randomness and the growth of prediction uncertainty,
let us consider the temporal evolution of the Pearson correlation coefficients 
between $\Delta X$ and $\Delta \mu_2$ and the standard deviation of $\Delta X$ plotted in Fig.~\ref{fig9}. 
The Pearson correlation coefficient measures the linear relationships between two variables, 
defined as the following equation
\begin{equation}
\rho=\frac{\operatorname{cov}(\Delta X, \Delta \mu_2)}{\sigma_{\Delta X} \sigma_{\Delta \mu_2}},
\end{equation}
where $\operatorname{cov}(\Delta X, \Delta \mu_2)$ stands for the covariance between  $\Delta X$ and $\Delta \mu_2$.
$\sigma_{\Delta X}$ and $\sigma_{\Delta \mu_2}$ stands for the standard deviation of $\Delta X$ and $\Delta \mu_2$, respectively. 
This measurement is between $0$ and $1$ to define the degree to which 
two variables move in coordination with each other.
For example, $\rho = 1$ denotes cases with perfect linearity while $\rho = 0$ denotes cases without any linear correlations.
We use it to demonstrate how the error function loses its linearity.

There also exist three zones with different characteristics, which loosely correspond to the proposed three areas as shown in Fig.~\ref{fig9} (a).
At first, $\rho = 1$, which belongs to the determinism stage when the linearity of the error function is perfectly maintained.
Then $\rho$ has an abrupt decline after some time, describing the loss of linearity between $\Delta X$ and $\Delta \mu_2$ during the transition stage.
In the end, $\rho$ falls below a negligible magnitude, and at the mean while $\Delta X$ and $\Delta \mu_2$ can be seen as unrelated random variables. 

Now, let us consider the history of the standard deviation of the horizontal prediction error $\sigma_{\Delta X}$.
As shown in Fig.~\ref{fig9} (b), $\sigma_{\Delta X}$ firstly grows exponentially along with time during the determinism and transition stages and then turns into a linear growth when entering the randomness stage. 
The linear growth results from that the plate reaches a statistically stable state in a chaotic attractor when the parametric noises have completely transferred into macroscopic randomness \cite{liao_numerical_2013}. 
Due to the exponential-linear growth law of $\sigma_{\Delta X}$, even extremely tiny physical perturbations can be quickly amplified to macroscopic scale and consequently result in a huge uncertainty in the long-term. 
Still take case 4 as an example.
The original error $\epsilon = 1\%$ for $\mu_2$ quickly climbs into a level that the maximum distance between the possible locations at $200$ UT reaches almost one hundred times larger than the length of the disk.

To quantify the predictability of the temporal invariant model, we further define $T_{ti}$, which measures the period of time that all trajectories do not have a noticeable divergence by $\sigma_{\Delta X} < 1$, as demonstrated in Fig.~\ref{fig9} (b).
There is also an exponential law between $T_{ti}$ and $\epsilon$ by the following formula:
\begin{equation}
T_{ti}=54.3071-17.9871\log_{10}{\varepsilon}
\label{eq:tti}
\end{equation}

\subsubsection{Temporal variant model}
We continue studying the history of the standard deviation of the horizontal prediction error by the temporal variant model. 
Physical perturbations are still set as $\epsilon = 1\%$, thus $\mu_2$ is a random variable that follows a normal distribution with a zero mean and the standard deviation $\sigma_{\mu_2}=0.002$.

The standard deviation of the horizontal prediction error also applies to the exponential-linear growth law as the temporal invariant model does. Thus, all conclusions still apply to the temporal variant model.

$T_{tv}$ has the same definition of $T_{ti}$, but is implemented in the temporal variant model. It is also an exponential law with regard to $\epsilon$ by the following formula:
\begin{equation}
T_{tv}=11.2357-19.8793\log_{10}{\varepsilon}
\label{eq:ttv}
\end{equation}

\begin{figure}[!t]
	\centering
	\includegraphics[scale=.4]{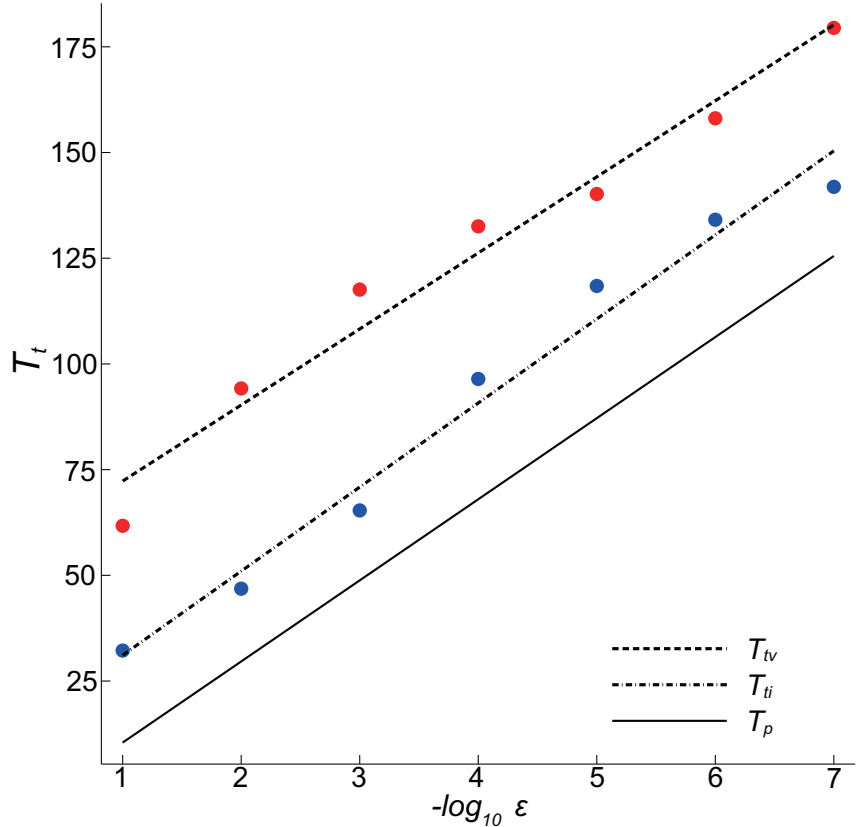}
	\caption{$T_p, T_{ti}$, and $T_{tv}$ against $\epsilon$. Red points: $T_{tv}$ data; blue points: $T_{ti}$ data; lines: fitted expressions.}
	\label{fig10}
\end{figure}

We directly compare formula (\ref{eq:tp}), (\ref{eq:tti}), (\ref{eq:ttv}) in terms of $T_p$, $T_{ti}$ and $T_{tv}$, shown in Fig.\ref{fig10}. 
It is obvious that $T_{tv}$ is longer than $T_{ti}$ when $\epsilon$ is at the same level. 
It is surprising that additional temporal randomness does sustain the predictability of the chaotic falling of plates. 
The results of the temporal variant model are more coherent with the experiments reported in the original work of APW model as repeatable\cite{andersen_analysis_2005, andersen_unsteady_2005}.
That is understandable since the APW model is only a finite-dimensional model extracted from the fluid-structure interaction problem of falling disks.
As discussed by Andersen et al., the model is successful due to the falling of plates are not strongly affected by its wake\cite{andersen_analysis_2005}.
Nevertheless, such wake would certainly introduce minute perturbations into the system and the true hydrodynamics shall fluctuate around the value predicted from the model.
That is a reason why temporal variant model is a more natural practice of introducing physical perturbations.
Also, it is promising to introduce data-driven models as amendments to the APW model to capture the physical perturbations and further extend the window of predictability in the future.

Another significant point lies in that all $T_p$, $T_{ti}$ and $T_{tv}$ have similar slopes.
That is coherent with our conclusions that the error growth rate is only related to the initial condition, since the error growth rate is determined by the manifold of the dynamical system.
Therefore, practically we can always compute the slope by $T_p$, and then obtain the slope from a single set of simulations or even experiments. By that, it is convenient to obtain a quantitative estimation of the predictability given the initial condition.

\subsubsection{Collective effects of physical perturbations}
\begin{figure}[!t]
	\centering
	\includegraphics[scale=.6]{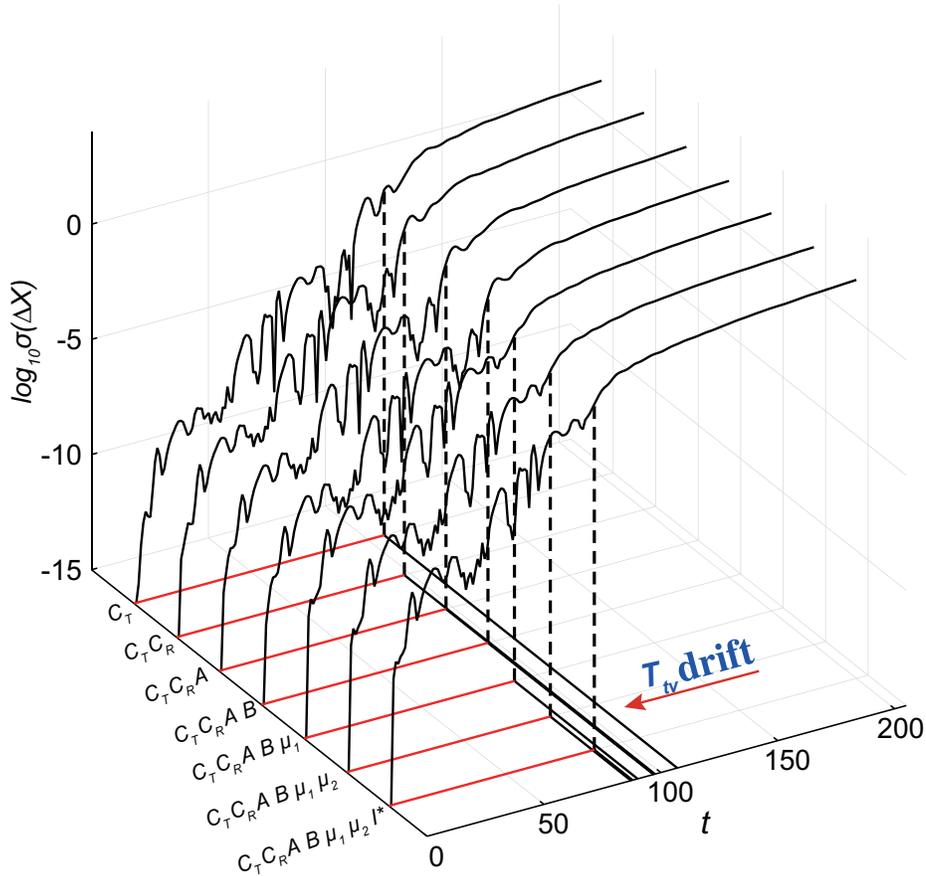}
	\caption{The temporal evolution of $\sigma_{\Delta X}$ with different parameters.}
	\label{fig11}
\end{figure}

In this part, we further study the collective effects of physical perturbations.
That is how $T_{tv}$ behaves under the influences of multiple physical perturbations contaminated by physical perturbations.
We gradually add the same fluctuation to all parameters in order of $C_T, C_R, A, B, \mu_1, \mu_2$, and $I^*$.
For example, the curve ``$C_T, C_R, A$'' in Fig.~\ref{fig11} contains noises in three parameters: $C_T, C_R, A$. 
As shown in Fig.~\ref{fig11}, the measurement $T_{tv}$ drifts backward when the number of parameters increases. 
To put it another way, the time to reach the same magnitude of the deviation (black horizontal dotted line in Fig.~\ref{fig11}) will be shortened.
That means different parameters have collective effects that will add to the overall evolution and thus accelerate the failure of reproduction.
It suggests that multiple parameters with physical perturbations will increase the rate of exponential growth.
It should be noted that the magnitudes of the physical perturbations are same but $T_{tv}$ changes irregularly. 
So different parameter combinations can change the value of $T_{tv}$.

\section{Replicability of experiments}

So far, we have studied the effects of physical perturbations on the predictability of the model. 
Naturally, the conclusions can be used to explain the replicability of experiments when there are unavoidable physical perturbations.
We use the the temporal variant model in this part since it suits the practical situations the best.
The quantitative relationships between $T_{tv}$ and $\epsilon$ are along with the dimensional analysis which was reported by Andersen et al\cite{andersen_analysis_2005} to reevaluate experiments.
The analysis can help us understand when physical perturbations do little harms in reality.


To directly compare with experiments, we first derive the non-dimensional variables of the experiments\cite{andersen_analysis_2005} as follows.
First, the magnitude of parameters in the APW model needs to be estimated.
These parameters can be divided to four main parts: circulation~($C_T$, $C_R$), drag~($A$, $B$), dissipation~($\mu_1$, $\mu_2$) and inertia~($I^*$). 
Higuchi\cite{higuchi_circulation_2001} provided the circulation of a disk through a series of experiments.
The fluctuation of the circulation around a disk can be obtained by fitting the experimental data with polynomials. 
So, the perturbations of $C_T$, $C_R$ are both $5.1\%$ if the contribution of $C_T$ and $C_R$ are the same.
According to Roos's drag experiments, the drag fluctuations are not significant and will never exceed $3\%$\cite{roos_some_1971}.
Therefore, the noises of $A$, $B$ are both around $3\%$ if the contribution of $A$ and $B$ are the same.
The dissipation terms are related to the viscosity, which depends on the ambient temperature.
Given that in most cases the falling process cannot last for a period 
that are long enough to see an obvious variation of the temperature, 
the perturbation of the dissipation is neglected here.
And for the same disk, the fluctuation of inertia is also zero.

\begin{figure}[!t]
	\centering
	\includegraphics[scale=.5]{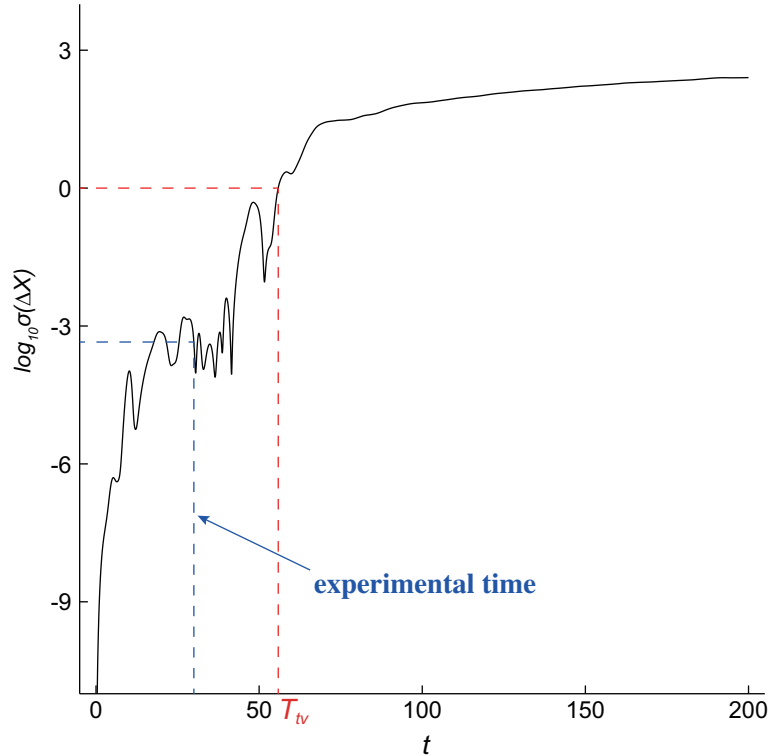}
	\caption{The temporal evolution of $\sigma_{\Delta X}$ in Andersen's experiments\cite{andersen_analysis_2005}.}
	\label{fig12}
\end{figure}

Then, we can explain why the chaotic trajectories in Andersen's experiments are repeatable.
We simulate the apparently chaotic trajectories to get $T_{tv}$ depending on the prior knowledge of physical perturbations.
After that, we transform the experimental time to the dimensionless time. 
By using dimensional analysis
\begin{equation}
	t^*=\frac{tU}{L},
	\label{eq:dim1}
\end{equation}
where 
\begin{equation}
	U=\frac{\nu Re}{L},
	\label{eq:dim2}
\end{equation}
Combine equation~(\ref{eq:dim1}) with equation~(\ref{eq:dim2}) and make $L=a$, we can obtain $t^*$ by the following relation
\begin{equation}
	t^*=\frac{\nu Re}{a^2}t,
\end{equation}
where $a$ represents the width of disk, $\nu$ is the viscosity of the water and $Re$ denotes the Reynolds number. 
The experimental time is $1.8s$\cite{andersen_analysis_2005} which gives the dimensionless time $30$ UT. 
As shown in Fig.~\ref{fig12}, $T_{tv}$ of Andersen's experiments is about $50$ UT. 
The dimensionless experimental time scale is way below the temporal criterion, 
and hence the noises have not been transferred to the macroscopic level.
This is why the experiments remain statistical stable and can be reproduced although they hold the chaotic nature.

\section{Conclusions}
Once was mentioned by Anderson et al.\cite{andersen_unsteady_2005} that 
``we note that Field et al. \cite{field_chaotic_1997} observed similar apparently chaotic motion for freely falling disks, 
whereas apparently chaotic motion was not found in the experiment by Belmonte et al. \cite{belmonte1998flutter}.''
And, to some extent this paradox is addressed in the present paper.
A temporal criterion is defined to measure how far a numerical prediction can go,
and simultaneously to tell if the chaotic falling is reproducible or not.
This study also provides a practical strategy to evaluate the credible prediction time 
by estimating the physical perturbations as a prior knowledge.

In the present work, we first examine the influence of known physical perturbations on the prediction of chaotic motion of free-fall disks.
As expected, the initial physical perturbations will lead to a false prediction of the falling trajectory.
In reality, the imprecise parameter is undetermined but a rough estimate of the statistic variables can be known from field tests.
We consider the temporal invariant and variant models to estimate the unknown physical perturbations that follows Gaussian distributions.
Monte-Carlo simulations are used to study the stochastic behavior.
Although errors of course turn larger as time passes, the exponential-linear growth pattern suggests that there is one temporal criterion to tell them apart.

While, experimentally, even chaotic falling can be repeated, 
which implies that the deviation of locations at least during the process is within an acceptable range of errors.
This phenomenon more or less contradicts the previous discussion.
By taking account of the fact that physical perturbations fluctuate at all time,
the outcome performs in a way that the temporal accumulation of noises can delay the linear growth 
and slow the exponential growth down.
This finding to a certain extent is consistent with the experimental observation.
It suggests that even if some system in nature may contain some chaotic subsystems, 
the system will keep stable because of the existence of physical fluctuations.
However, if we simplify a complex system like turbulence, it has a great chance that some of the crucial parameters are ignored,
so, the simplified system like Lorenz system will be totally unstable, sensitive and chaotic.

A promising way to lengthen the prediction time of chaotic problem when there exist physical perturbations is the data assimilation,
which embeds a small portion of observed data during the numerical simulation to introduce amendments to the computation.
This practical method is the future direction to approach the long-term prediction when physical perturbations co-exist with chaos.

 \section*{Acknowledgments}

This work is supported by the National Natural Science Foundation of China (No. 91752104).



\bibliographystyle{unsrt}

\end{document}